\documentclass[11pt,a4paper]{article} 
\usepackage{graphicx} 
\usepackage{amsfonts}
\begin{document} 
\thispagestyle{empty}
\newcommand{\hs{\hspace{20pt}}}
\small\normalsize
\frenchspacing

\noindent
{\Large\bf{A conjecture concerning  determinism, reduction, and\\[3pt]measurement in quantum mechanics}}\hspace{1.5pt}\footnote{An earlier version of this article was paper-published in Quantum Studies: Mathematics and Foundations {\bf3} (4) 279-292 (2016). The content of the present version is the same but the presentation is improved.}
\\
\\ 
{\bf{Arthur Jabs}}
\renewcommand{\baselinestretch}{1}
\small\normalsize
\\
\\
Alumnus, Technical University Berlin. 

\noindent
Vo\ss str. 9, 10117 Berlin, Germany
\\
\noindent
arthur.jabs@alumni.tu-berlin.de
\\						
\\
(24 July 2019)
\newcommand{\rmi}{\mathrm{i}}
\vspace{15pt}

\noindent
{\bf{Abstract.}} Determinism is established in quantum mechanics by tracing the probabilities in the Born rules back to the absolute (overall) phase constants of the wave functions and recognizing these phase constants as  pseudorandom numbers. The reduction process (collapse) is independent of measurement. It occurs when two wavepackets  overlap in ordinary space and satisfy a certain criterion, which depends  on the phase constants of both wavepackets. Reduction means contraction of the wavepackets to the place of overlap. The measurement apparatus fans out the incoming wavepacket into spatially separated eigenpackets of the chosen observable. When one of these eigenpackets together with a wavepacket located in the apparatus satisfy the criterion, the reduction associates the place of contraction with an eigenvalue of the observable. The theory is nonlocal and contextual.

\vspace{5pt}
\begin{list}
{\textbf{Keywords:}}
{\setlength{\labelwidth}{2.0cm}
 \setlength{\leftmargin}{2.2cm}
 \setlength{\labelsep}{0.2cm} }
 \item
determinism, overall phases, hidden variables, reduction, collapse,  localization, quantum measurement, Born rule
\end{list}
\vspace{-4pt}
\vspace{15pt}
\\
\rule{\textwidth}{.3pt}
\newcommand{\leer}{\hspace*{\fill}}
\newcommand{\lll}{\hspace*{10pt}}
\newcommand{\hsp}{\hspace*{60pt}}
\begin{center}
\vspace{-20pt}
\noindent
\hsp 1~~ Introduction \leer 2\hsp \lll \\ 
\hsp 2~~ The absolute phase constants \leer 3\hsp \lll \\
\hsp 3~~ Reduction \leer 6\hsp \lll \\
\hsp 4~~ The spacetime nature of measurements \leer 8\hsp \lll \\ 
	\hspace{20pt}\hsp 4.1~~Fanning out, 4.2~~Contraction,   \leer 8\hsp \lll \\      
	 \hspace{20pt}\hsp 4.3~~Dynamical interaction, 4.4~~Magnification \leer 8\hsp \lll \\ 
\hsp 5~~ Reproducing the Born rules \leer 12\hsp \lll \\ 
\hsp Appendix A. Valuation of the absolute phase \leer 16\hsp \lll \\
\hsp Appendix B. Approximation in Eq.~(5.4) \leer 17\hsp \lll \\
\hsp Notes and references	\leer 18-21\hsp \lll \\
\end{center}
\vspace{-10pt}
\rule{\textwidth}{.3pt}
\\

\newcommand{\rmf}{\mathrm{f}}
\newcommand{\rmd}{\mathrm{d}} 
\newcommand{\sbl}{\hspace{1pt}}
\newcommand{\bfitp}{\emph{\boldmath $p$}}
\newcommand{\bfitr}{\emph{\boldmath $r$}}
\newcommand{\bfitsr}{\emph{\footnotesize{\boldmath $r$}}}
\newcommand{\bfitQ}{\emph{\boldmath $Q$}}
\newcommand{\bfitk}{\emph{\boldmath $k$}}
\newcommand{\PSI}[1]{\Psi_{\textrm{\footnotesize{#1}}}}
\newcommand{\PHI}[1]{\Phi_{\textrm{\footnotesize{#1}}}}
\newcommand{\spsi}[1]{\psi_{\textrm{\footnotesize{#1}}}}
\newcommand{\pcop}{P_{\textrm{\footnotesize{Cop}}}}
\newcommand{\pred}{P_{\textrm{\footnotesize{red}}}}
\newcommand{\psis}{\psi_{\rm s}(\bfitr,t)}
\newpage
\begin{flushright}
\emph{Citius emergit veritas ex errore\\
quam ex confusione} 

Francis Bacon
\end{flushright}

\bigskip

\noindent
{\textbf{1~~Introduction}}
\smallskip

\noindent
The probabilities in present-day quantum mechanics are fundamentally different from those in classical (pre-quantum) statistical mechanics. In statistical mechanics the probabilities are conceived to be reducible to finer details of the physical situation considered. In an ideal gas, for example,  the finer details are the positions and momenta of the individual particles. In principle, classical statistical mechanics is a deterministic theory. Any random appearance of macroscopic quantities is due to their very sensitive dependence on the microscopic initial conditions \cite{Smoluchowski}, \cite{Wolfram}.

In contrast to this, the probabilities in quantum mechanics, according to contemporary orthodoxy, are irreducible to any underlying more detailed specification. According to this view, quantum mechanics is irreparably indeterministic. As is well known, this was always considered a serious drawback by Einstein, and Dirac wrote \cite{Dirac73}:
\begin{quote}
It may be that in some future development we shall be able to return to determinism, but only at the expense of giving up something else, some other prejudice which we hold to very strongly at the present time.
\end{quote}

The general procedure of making quantum mechanics deterministic is to introduce additional variables that in principle determine the outcome of each individual measurement, and over many repetitions satisfy the approved Born probability formulas. Such variables are usually called hidden variables or hidden parameters, terms originally coined by v.~Neumann \cite{Neumann}.
Actually, some of the parameters or variables that have been considered in the literature are not hidden at all, so terms like `uncontrolled' \cite[p.~92]{Bell87}, `determining' or `fixing' variables would be more appropriate. Nevertheless, following entrenched usage, we also speak of hidden variables. 

In the present approach hidden variables are introduced, which are equated with the absolute (global, overall, spacetime-independent) phase constants of the quantum mechanical wave functions. Each wave function which represents an individual quantum object is conceived to contain an individual phase factor $\exp(\rmi \alpha)$ with a constant phase $\alpha$. In an ensemble of objects the phase constants $\alpha$ are thought of as being random numbers uniformly distributed in $[0,\,2\pi]$. They are, however, conceived to be \emph{pseudo}random numbers. That is, the phase constants only seem to be random, but in reality they are determined by certain initial conditions. This is in the spirit of the theory of deterministic chaos, which has been systematically developed since the 1960s \cite[p.~971]{Wolfram}. 

The experimentally confirmed violations of the Bell inequality show that no hidden variables can exist that would lead to a local description of nature. Another restriction comes from the Bell \cite{Bell87} and the Kochen-Specker \cite{Kochen67} theorems and requires any hidden-variable theory to be `contextual' if it is to reproduce the predictions of standard quantum mechanics. In the case of hidden-variable models, contextuality means that the outcome of an experiment depends upon hidden variables in the apparatus \cite{Conway}. This is the case in the present conjecture. The Bell inequalities are violated because our phase constants cannot be taken as common causes in the past, those in different apparatuses being independent of each other. Actually, whatever no-go theorems are brought forward, the present approach does reproduce the Born rules and with them can violate the Bell inequalities. This means nonlocality.

In Sec.~2 we introduce the absolute phase constants of the wavepackets and point out experiments where they can be determined and what their role is in superpositions, transformations and in the reduction process.

In Sec.~3 a criterion is formulated, which decides, independent of any measurement, when and where a reduction will occur. The reduction is conceived to be a spatial contraction. The crucial new ingredient is that the phase constants of two wavepackets are involved.

Sec.~4 introduces a new view on measurement: when the incoming wave function is developed into eigenfunctions of the observable, the apparatus achieves that the eigenfunctions occupy separated regions of space and a reduction associates an eigenvalue with the position of an observable spot. 

Sec.~5 then describes how and under what approximations the criterion reproduces the Born probability rules.

The Appendices justify some more technical assumptions.

\vspace{10pt}
\noindent
{\textbf{2~~The absolute phase constants}}
\nopagebreak
\smallskip

\noindent
The absolute phase constants, which  in the present approach are the physical quantities that play the role of the pseudorandom numbers, are nonlocal hidden variables. Actually, they have already been proposed  by Ax and Kochen \cite{Ax}, where the authors state in the abstract that:  
\begin{quote}
In the new interpretation, rays in Hilbert space correspond to ensembles, while unit vectors in a ray correspond to individual members of such an ensemble. The apparent indeterminism of SQM [statistical quantum mechanics] is thus attributable to the effectively random distribution of initial phases. 
\end{quote}

Ax and Kochen's elaboration of the idea has a strong mathematical orientation. It is not conceived to be deterministic and differs in several other respects from the rather physically oriented elaboration in the present article. But evidently Ax and Kochen consider the equating of the phases with the hidden variables not to be \emph{a priori} forbidden by the Kochen-Specker contextuality theorem \cite{Kochen67}.

Another hidden variable, conceived to be deterministic, is the initial position of the point particle in the de Broglie-Bohm theory. In itself it is obviously a local variable. The nonlocal character of the de Broglie-Bohm theory is based on the existence of spatially separated entangled wave functions in $3N$-dimensional configuration space, with the same time variable for all $N$ particles. This theory has not found broad acceptance among physicists. I suppose that one of the deepest reasons for this is that its basic element is a point particle. Einstein, for example, wrote \cite{Einstein34}:
\begin{quote}
On the other hand, it seems to me certain that we have to give up the notion of an absolute localization of the particles in a theoretical model. This seems to me to be the correct theoretical interpretation of Heisenberg's indeterminacy relation.
\end{quote}

In the present theory, the concept of a point particle is replaced by the concept of a real, objective wavepacket  of finite extension, according to the realist interpretation expounded in \cite{Jabs92} - \cite{Jabs06}. The argument $\bfitr$ in the wave function here does not mean the position of a point particle. Rather, the position of the wavepacket is given by the (not explicitly shown) parameter $\bfitr_0$ in $\psi(\bfitr,t)$, which specifies its center given by  $\int\bfitr|\psi(\bfitr,t)|^2\rmd^3r$, say.  Such a wavepacket is endowed with 
special properties implying reductions and correlations between results of spacelike separated measurements that are not fully determined by causes in the common past. We argue with Schr\"odinger wavepackets, but the solutions of any of the quantum mechanical equations of motion, and even photon packets of electromagnetic waves, do as well. A more detailed description of that interpretation is not needed here, but one consequence is that the concept of a point position is never used in the present article, and when we speak of a particle we always mean an extended wavepacket. Readers who are still determined adherents of the Copenhagen interpretation can easily translate our formulations into the language they prefer.

Now, the fact that in the Born probability rule
\begin{displaymath}
p\propto |\psi(\bfitr,t)|^2
\end{displaymath}
the phase of $\psi$ does not appear at all, immediately raises the question whether this might not be the very reason for the probabilistic feature. Actually, Born raised that question already in the first two papers in which he proposed the probability interpretation \cite[p.~826, 827]{Born26b},  \cite[p.~866]{Born26a}, although only briefly and without pursuing the matter further. In \cite[p.~866]{Born26a} he wrote: 
\begin{quote}
... we have so far no reason to believe that there are some inner properties of the atom which condition a definite outcome for the collision. Ought we to hope later to discover such properties (like phases of the internal atomic motion) and determine them in individual cases?
\end{quote}

We thus conceive each wave function of the standard formalism of quantum mechanics which represents a quantum object to contain an individual phase factor $\exp(\rmi \alpha)$, with $\alpha$ being a real number, independent of $\bfitr$ and $t$. The wave functions are still normalized to 1. In an ensemble of such wave functions the $\alpha$\sbl s are pseudorandom numbers. They are assumed to be uniformly distributed in the interval $[0,\, 2\pi]$ if they
refer to some kind of equilibrium, that is, if we do not select sets of wave functions with determined phase values and then isolate them (in those situations where it is possible; see below). This is a postulate. It fits with the postulate of random phases already met in quantum statistical mechanics \cite[p. 173, 190]{Huang}. And it is analogous to the postulate in classical statistical mechanics that in equilibrium the density of points in phase space is uniform \cite[p.~129]{Huang}. In other words, probabilities in quantum mechanics will be traced back to a uniform distribution of phases in the same way as probabilities in classical statistical mechanics are traced back to a uniform distribution of points in phase space.

The total wave function written in polar form reads
\begin{displaymath}
\psi\!(\bfitr,t)=e^{\rmi \alpha}\,e^{\rmi \varphi\!(\footnotesize{\bfitr},t)}|\psi\!(\bfitr,t)|= e^{\rmi(\alpha+\varphi\!(\footnotesize{\bfitr},t))}|\psi\!(\bfitr,t)|,
\end{displaymath}
where  $\;\exp(\rmi\varphi\!(\bfitr,t))\,|\psi\!(\bfitr,t)|\;$ is a normalized solution of the Schr\"odinger equation, $\alpha+\varphi\!(\bfitr,t)$ is the total phase, and $\alpha$ is the absolute phase constant. If $\alpha$ is uniformly distributed in $[0,\,2\pi]$, then $\alpha+\varphi\!(\bfitr,t)$
modulo $2\pi$ for any fixed $\bfitr$ and $t$ is also so distributed \cite{Feller}. Thus, with respect to random appearance the total phase and the phase constant play the same role, and it is irrelevant which moment we choose as the initial moment for fixing the phase constant.
  
The absolute phase constants are physical in the sense that physical situations exist where they can be determined. For example, when two independent weak quasi-monochromatic laser beams (photon wave functions) are superposed \cite{Paul04}, \cite{Kaltenbaek}. In the superpositions the absolute phases become relative phases and determine the positions of the interference fringes. This can also be said of the wave functions of Bose-Einstein condensates of atoms \cite{Laloe} and photons \cite{Klaers}. Note that it is not possible to speak of a definite phase difference between two independent wave functions if these functions do not each have a definite phase of their own (\cite[p.~41]{Ax}).

\vspace{10pt}
\noindent
The following properties are attributed to the phase constants:

What happens  when a one-particle wavepacket $\psi$ is written as $\psi=\sum_i \lambda_i \phi_i$? The $\phi_i$\sbl s  here are simply terms in a largely arbitrary mathematical decomposition and no single   $\phi_i$  represents the whole particle. As the phase constant refers to the wavepacket that represents the whole particle, it is put in front of $\psi$ and so in front of each of the $\phi_i${\sbl}s. If the $\phi_i$\sbl s were to contain individual random phase factors $\exp(\rmi \alpha_i)$, then, after averaging over the \mbox{$\alpha_i$\sbl s}, there would be no interference terms in the probability expressions, and the superposition could only be a statistical mixture (cf. \cite[p.~254]{Cohen}).

When two independent wavepackets $e^{\rm i\alpha_1}\psi_1$ and $e^{\rm i\alpha_2}\psi_2$  get entangled, the phase constant of the wave function of the entangled system is ($\alpha_1+\alpha_2$) modulo $2\pi$.

\vspace{5pt}
What happens to the absolute phase constants when the unitary Hilbert space  operators $U$ as representations of the symmetry groups act on the wave functions? In the traditional ray representation of the wave functions, where their absolute phases are given no physical significance, the operators also need only to be ray (projective) representations, that is,   $\;U_iU_j=e^{\rmi\alpha(i,j)}\,U_k$,   with real $\alpha(i,j)$.
However, this means that by a mere transformation, the wave function may acquire an additional arbitrary contribution to its absolute phase constant.
As our theory is based on a vector representation of the wave functions, where their absolute phases do have physical significance, we must also insist on a vector  (usual) representation of the transformation operators  \cite{Feynman}, \cite{Sexl}, so that $U_i       U_j=U_k$.

Although different wavepackets (particles) have different absolute phase constants, these, like the different shapes, do not play any role in the usual treatment of `identical' particles. The only exception is when reduction occurs. Indeed, the absolute phase constants become effective exclusively in the reduction processes. What happens with the phase constants in the reduction process is discussed in Sec.~3. 
                                                                                          
\vspace{10pt}
\noindent
{\textbf{3~~Reduction}} 
\smallskip

\noindent
In the present theory the reduction is a real physical process which has nothing to do with measurement. Therefore,  we describe reduction  before we come to the role which it plays in a measurement. The description will not be in (necessarily nonlinear) analytical terms complementing the Schr\"{o}dinger equation but in terms of a  criterion specifying when and where the Schr\"{o}dinger evolution is interrupted by a reduction. In this we are guided by known facts.

Thus we conjecture:

(a) A reduction can occur when two wavepackets  overlap, of which at least one represents a massive elementary particle or a cluster of them. 

A  cluster consists of several elementary particles bound together, such as an atom, a molecule, or a larger compound. The wavepacket $\Psi$ representing such a (free) cluster is the product of two wave functions
\begin{displaymath}
\hspace{105pt} \Psi=e^{\rmi \alpha}\,\psi(\bfitr,t)\times \psi_{\rm R}(\rho_1,\cdots,\rho_N,t)\hspace{100pt}(3.1)
\end{displaymath}
namely a \emph{center-of-mass (CM) function} $e^{\rmi\alpha}\psi(\bfitr,t)$, which is a superposition of de Broglie waves of a particle with the mass of the cluster as a whole, and of an \emph{internal function} $\psi_{\rm R}(\rho_1,\cdots,\rho_N,t)$, which represents the relative positions $\rho_i$ of the constituent elementary particles \cite[Ch. IX, \S 12, 13]{Messiah}. In the remainder of this article $e^{\rmi\alpha}\psi(\bfitr,t)$ will mean the CM wave function. For an elementary particle there is no quantum mechanical internal wave function in the present sense; its wave function is counted among the CM functions.

(b) The reduction is a sudden \emph{spatial contraction} of both CM wave functions involved. 
Each of the two functions contracts to a volume of the order of the volume $\upsilon_{\mbox{o}}$ of the spatial overlap region, i.e. where $E:= |\spsi{1}(\bfitr,t)| \, |\spsi{2}(\bfitr,t)|$ is practically concentrated (its effective support) at the first time the criterion (3.2) given below is satisfied; for example the region where $E$ then is larger than $1/e^2$ of its maximum value. The center of the contraction is around the center of that overlap volume. Of course,  $\Delta x\Delta y\Delta z\times \Delta p_x\Delta p_y\Delta p_z \ge (\hbar/2)^3$ must be satisfied.

Note that the spatial extension of the CM wave function $e^{\rmi\alpha} \psi(\bfitr,t)$ can be very much larger than the de Broglie wavelength $\lambda_{\rm dB}=\hbar/m_0v$, where $m_0$ is the mass and $v$ the velocity of the cluster as a whole, and it can also be very much larger than the extension of the internal wave function. This is in accordance with the interference patterns observed with atoms and complex molecules at diffraction gratings \cite{Hornberger} - \cite{Arndt10}. These experiments show that it is the CM function of the atom or molecule that is responsible for the interference patterns. Numerical estimates show that the initial volume of this function, supposing it is equal to the extension of the internal wave function, is small compared with the slit separation, and thus cannot produce interference effects. But due to spreading on the way from formation through vacuum to the grating (cf. e.g. [11, Appendix A]), its extension becomes larger than the slit separation and then does produce the interference pattern.

(c) The spatial form of the contracted wave functions is Gaussian (in $x, y,z$). This is the form of the minimum wave function in the sense of the Heisenberg relations with equality sign. The phase constant is not changed.

(d) If one of the two overlapping wavepackets represents an entangled system of two wavepackets, its phase constant, as stated earlier, is the sum of the phase constants of the constituent particles modulo $2\pi$, and this phase constant is the one that enters  the formula (3.2a) as $\alpha_1$ or $\alpha_2$, respectively. When the two wavepackets which represent the entangled system are spatially well separated, as e.g. in the Einstein-Podolsky-Rosen situation, the reduction contracts these wavepackets around corresponding points.  Moreover, reduction disentangles these two, and we stipulate that each one takes over the phase constant of the entangled system.

\vspace{8pt}
Now, the criterion for such a contraction to occur is conjectured to consist of the two conditions:

\begin{displaymath} 
\hspace{150pt}
 |\alpha_1-\alpha_2| \leq \mbox{{\normalsize{$\frac{1}{2}$}}}\,\alpha_{\textrm s} \hspace{140pt}(3.2\rm a)
\end{displaymath} 

\vspace{-10pt}

\[
\hspace{90pt} 
K:= \bigg[\int_{\mathbb{R}^3} |\spsi{1}(\bfitr,t)| \, |\spsi{2}(\bfitr,t)| \, {\textrm d}^3r\bigg]^2 \ge \alpha/{2\pi}.
\hspace{68pt}(3.2\rm b)
\]

\vspace{12pt}

Formula (3.2a) is a `phase-matching condition'. $\alpha_{\textrm s}=e^2/\hbar c \approx 1/137$ is Sommerfeld's fine structure constant. The  CM wave functions  \sbl $\exp(\rmi\alpha_1)$$\spsi{1}(\bfitr,t)$ and   $\exp(\rmi\alpha_2)$$\spsi{2}(\bfitr,t)$   become capable of contraction when one phase constant lies in an interval of size  $\alpha_{\textrm s} $  around the other. 

Formula (3.2b) is an `overlap condition'. The phase constant $\alpha$ is the smaller one of $\alpha_1$ and $\alpha_2$. As the functions are normalized, the quantity $K$ lies in the interval $[0,1]$.

The functions $\exp(\rmi \alpha_1)\spsi{1}$ and $\exp(\rmi \alpha_2)\spsi{2}$ may pass by each other only at some distance and their overlap may be accomplished only by their respective marginal regions.  In any case the integral depends on time. The moment of contraction is as soon as (3.2b) is satisfied, given that the wave function has already encountered a phase matching cluster, i.e. (3.2a) is  already satisfied.

The contraction is sudden, that is, it may occur with superluminal speed. Experimental evidence, existing since 1913, is collected in \cite[Secs.~2.3, 3.1]{Jabs96}, \cite{Jabs15}. It is one aspect of the by now well known quantum mechanical nonlocality.
\vspace{4pt}

\noindent 
Some remarks may be of interest:

1) The conjecture goes beyond a re-interpretation of the standard quantum mechanical formalism. Formulas (3.2) are on an equal footing with the Schr\"odinger equation in describing the temporal evolution of the wavepackets.

2) The formulas are symmetric in the wavepackets.

3) The phase-matching interval is conjectured to be of the order of Sommerfeld's fine structure constant $\alpha_{\textrm s}$ because the phases are dimensionless and $\alpha_{\textrm s}$ is the only dimensionless fundamental constant of nature which plays a role in quantum mechanics (hydrogen-like atoms \cite{Bethe}, natural linewidth \cite{Jauch}). 

4) The contraction in our theory is a \emph{localization}, but a more radical one than that in the decoherence theories  \cite{Adler} - \cite{Joos07}, where localization only means that through creation of entanglement between a quantum system and its (quantum) environment, `phase relations between macroscopically different positions are destroyed' \cite[p.~1, 6]{Joos07}, \cite{note2}. Our contraction is responsible for the persistent illusion of a point particle.

5) Contraction here is not the weird process of reduction directly to the eigenfunction of some arbitrarily chosen observable. `Observable' is suggestive of `measurement', but measurement should not even be mentioned here.

6) The appearance of the signs $\le$ and $\ge$ in the criterion (3.2) means that several initial conditions can lead to contraction. But afterwards they leave no trace, so there is no longer time reversal symmetry in reduction/contraction.

7) Formulas (3.2) are relations between dimensionless numbers, and so their form is Lorentz-invariant. 

8) The theory also explains why macroscopic objects show no wave properties \cite{ Jabs19}.

9) Though the absolute phases and the reduction process open a door to a deterministic quantum mechanics,  the details of the specifications here proposed may be modified in future  elaborations. The case of three or more overlapping wavepackets, for example, being far less frequent, is not dealt with here. 

\vspace{5pt}

In the remainder of this article we will no longer show the phase factors $\exp(\rmi\alpha)$ explicitly, but will regard them again as hidden in the symbols $\psi$.

\vspace{10pt}
\noindent
{\textbf{4~~The spacetime nature of measurements}}
\smallskip

\noindent
Our conception  of a quantum mechanical measurement is different in several respects from most other treatments. There is an abundance of literature on the measurement process in quantum mechanics \cite{Wheeler} and it is not our purpose to review it here. We restrict ourselves to those aspects that are relevant to our approach.
One point of difference is that we take the stand that it is \emph{always} possible to relate a property of a system by deterministic physical laws to spacetime events, so that the measurement of any property is ultimately explained by one or several spacetime position measurements. This is not a new idea. Einstein, for one, wrote \cite{Einstein70}:
\begin{quote}
Now it is characteristic of thought in physics, as of thought in natural science generally, that it endeavours in principle to make do with ``space-type'' concepts \emph{alone}, and strives to express with their aid all relations having the form of laws. The physicist seeks to reduce colours and tones to vibrations, the physiologist thought and pain to nerve processes, in such a way that the psychical element as such is eliminated from the causal nexus of existence, and thus nowhere occurs as an independent link in the causal associations.
\end{quote}
And similar statements have been made by Bell \cite[p. 10, 34, 166]{Bell87} and many others, for example \cite{Holland93},  \cite{Sussmann}. In the cited papers the idea is not further elaborated. In our treatment it plays a fundamental role. Thus, the measurements we are concerned with are typical quantum mechanical measurements, that is those where the extended wavepacket nature of the particles plays a role. This implies that the spatial resolving  interval of the measuring apparatus is smaller than the spatial extension (volume) of the wavepacket  representing the measured object. Thus, the measurement proceeds in four steps: fanning out, contraction, dynamical interaction, and magnification.

\vspace{5pt} 
\noindent
{\bf{4.1\emph{~Fanning out}}}

\noindent
In the first step
the \emph{active region} of the measuring apparatus must achieve that the    \begin{figure}[h]
\begin{center}
\includegraphics[width=0.7\textwidth]{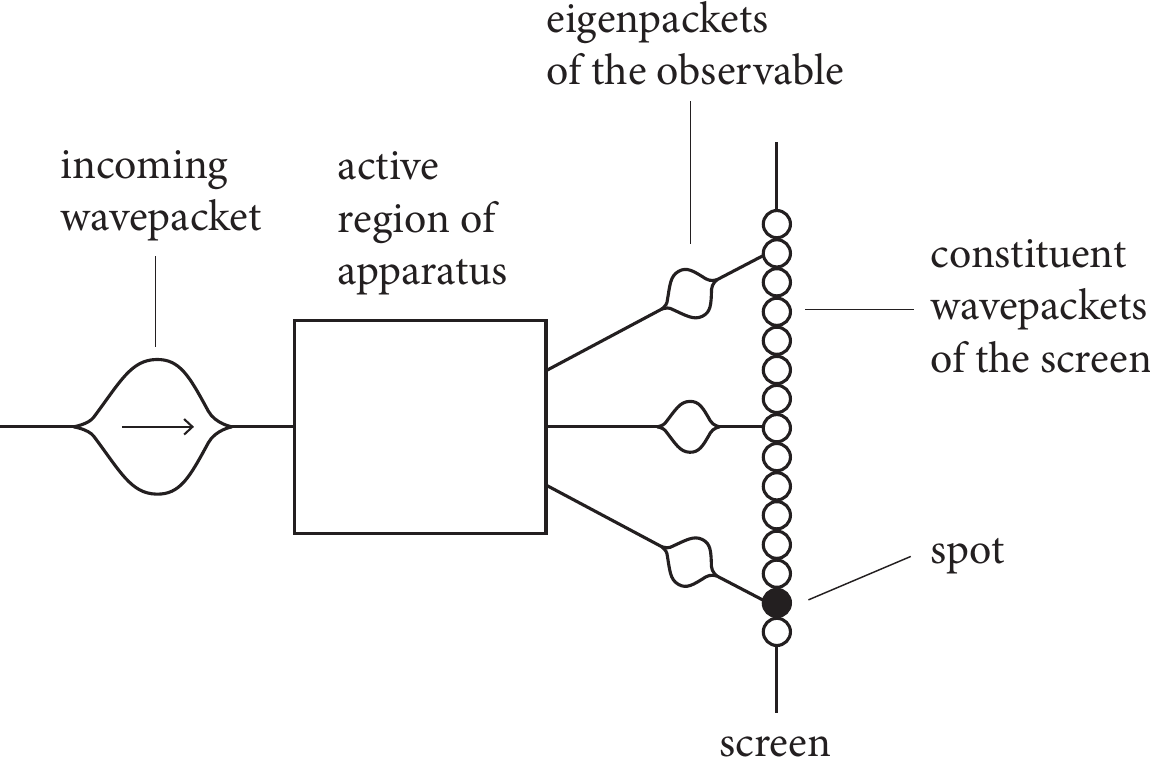}
\caption{Sketch of the apparatus fanning out the incoming wavepacket}
\end{center}
\end{figure} 
 
\noindent
incoming wavepacket $\psi_{1}(\bfitr,t)$, develops into a superposition
\[
\hspace{150pt} \psi_{\rm s}(\bfitr,t)=\sum_l c_l\,\psi_l(\bfitr,t) \hspace{115pt}(4.1)
\]
\vspace{-10pt}
\[
\textrm{with}\hspace{79pt} c_l=(\psi_l(\bfitr,t),\psi_{\rm s}(\bfitr,t))=\int\psi_l^*(\bfitr,t)\,\psi_{\rm s}(\bfitr,t)\,\rmd^3r \hspace{55pt}(4.2)
\]
or, in the continuous case,
\vspace{-10pt}
\[
\hspace{140pt}\psi_{\rm s}(\bfitr,t)=\int c(a)\,\psi(a;\bfitr,t)\rmd a \hspace{100pt}(4.3)
\]
\vspace{-10pt}
\[
\textrm{with}\hspace{50pt} c(a)=(\psi(a;\bfitr,t),\psi_{\rm s}(\bfitr,t))=\int\psi^*(a;\bfitr,t)\,\psi_{\rm s}(\bfitr,t)\,\rmd^3r \hspace{55pt}(4.4)
\]
with normalized eigenpackets $\psi_l(\bfitr,t)$ and $\psi(a;\bfitr,t)$, respectively, of the observable (self-adjoint operator) corresponding to the physical quantity under consideration, 
\vspace{8pt}

\emph{subject however to the condition that the different eigenpackets are located in} 

\emph{different regions of space}. 
\vspace{8pt}

\noindent
A fresh look at what is actually done in real experiments shows that every apparatus fans out the incoming wavepacket into a coherent superposition of eigenpackets of the observable, which are located  in separate regions of space. This \emph{fanning-out} step is an essential new ingredient of our theory.  It associates a particular region of space with a particular eigenvalue of the observable. It is not a mathematical but a physical problem. The mathematical expansions (4.1), (4.3) are always possible. But the wavepacket must be exposed to such a physical situation in which the eigenfunctions \{$\psi_l$\} and \{$\psi(a)$\}, respectively, become spatially separated from each other. The design of an apparatus with the physical laws accomplishing this is a challenge to the inventiveness and ingenuity of the experimental physicist. The most obvious examples are the prisms, polarizing beam splitters, diffraction gratings, and the magnetic field in the Stern-Gerlach device.

\vspace{5pt}
\noindent
{\bf{4.2\emph{~Contraction}}}\\
According to what was stated in Section 2, all superposed eigenpackets $\psi_l(\bfitr,t)$ or $\psi(a;\bfitr,t)$  have the same absolute phase constant as the original incoming wavepaket. They are all coherent and together represent the incoming particle. They then enter the \emph{sensitive region} of the apparatus. We call it the \emph{screen} (e.g. a photographic layer). This is the region (`context') where the contractions are to occur. The screen consists of many very small incoherent cluster wavepackets with differing phase constants. There is certainly one cluster whose phase constant matches the one of the eigenpackets and satisfies the contraction condition (3.2b), where $\psi_1$ now is the incoming wavepacket and $\psi_2$ one of the cluster packets of the screen. The overlap volume between these two is very small because the cluster packets are very small. The incoming wavepacket is thus effectively contracted to the place of one of the cluster packets.

That different cluster wavepackets have different phase constants is a crucial condition in our approach. The facts that the clusters are very small, and that in calculating scattering amplitudes it is an approved approximation that each scattering center in the target material acts as if it were alone \cite[p. 370]{Messiah}, favors our assumption. 

The incoming wavepacket, according to Section~3, turns into a Gauss function of the order of the overlap volume $\upsilon_{\mbox{o}}$, that here is practically the volume of the clusters in the screen or part of it. It can be assumed that the bounds which keep the cluster together are much stronger than those between the cluster and its environment, so that in first approximation Eq.~(3.1) applies here too. In any case the incoming wavepacket after the contraction occupies the place of a particular eigenfunction of the observable. This wavepacket need not be exactly equal to the eigenpacket but can be taken to be a good approximation to it.

\vspace{5pt}
\noindent
{\bf{4.3\emph{~Dynamical interaction}}}\\
The incoming wavepacket after contraction interacts with the wavepacket that represents the cluster.  This is a dynamical interaction governed by the Schr\"odinger equation. It leads to the final wavepacket $\spsi{f}(\bfitr,t)$, and the internal wave functions have participated in this.

The final wavepacket $\spsi{f}$ may  be absorbed (photons, neutrons, atoms), or it may escape the interaction region. If it escapes, it may be close to the contracted packet $\spsi{c}$ or it may differ appreciably from it, depending on the details of the interaction. As $\spsi{c}$ may not be exactly equal to the eigenpacket $\psi_{n}$ of the observable, so also $\spsi{f}$ may not be exactly equal to $\psi_{n}$. It is, however, reasonable to assume that $\psi_{\rmf}$ immediately after the interaction is also, like $\spsi{c}$, concentrated about the position $\bfitr_0$ of the cluster. Therefore, though we cannot generally say that when the measurement is over, the incoming superposition of eigenpackets has turned into a function that is exactly equal to an eigenpacket $\psi_n$ of the observable (the traditional postulate), we may suppose that usually this happens to an acceptable degree of approximation. The exact degree of approximation is difficult to determine, because there are many thought experiments about the wavepacket $\psi$ when it, after contraction, leaves the measuring apparatus, but to my knowledge there is no real experiment concerned with this question.

Photons are a special case. In contrast to massive particles they vanish completely in the act of reduction, and speaking of the place where a vanished thing \emph{is} does not seem to make much sense. However, the statement that a one-photon wavepacket, due to its quantum nature, is absorbed as a whole does not in itself imply that the absorption occurs at one narrow place, narrower than the wavepacket's original extension. As this is indeed the case, we continue to speak of contraction for both massive and massless particles.

Whether $\spsi{f}$ will get entangled with other packets is not relevant because this does not change the result of the measurement.

The first three steps may thus be summarized as: $\spsi{1}\rightarrow\spsi{s}\rightarrow\spsi{c}\rightarrow\spsi{f}$, where
 \\$\spsi{1}=  $ incoming wavepacket, $\spsi{s}$ = superposition of spatially separated 
eigenpackets
\\$\psi_n$, \hspace{2pt} 
 $\spsi{c}$ = contracted wavepacket,  and 
 $\spsi{f}$ = escaping final wavepacket.

\vspace{5pt}
\noindent
{\bf{4.4\emph{~Magnification}}}
\nopagebreak

\noindent
The dynamical interaction of the contracted wavepacket  with the cluster wavepacket must lead to a magnification, that is, to a macroscopic,  observable \emph{spot} in the screen. A typical case of spot formation is an atom which is ionized, excited or de-excited and from which an avalanche develops, involving more and more particles (`secondary electron multiplier').

Not all clusters are capable of spot formation, those that are, we call `sensitive' clusters or \emph{grains} (\emph{grain wavepackets}). As one example think of the grains in a photographic emulsion. Generally, the magnification still involves a microscope or a chain of apparatuses with a great deal of electronic equipment. The spot formation is the step that accomplishes the ``circumstancial evidence'' in the macroworld of what happened in the microworld. But the details of this process are beyond the intended scope of the present article. The observation of the spot then is like any classical observation of what already \emph{is} and concludes the measurement.

The dependence of the criterion on the phases of both the incoming particle ($\alpha_1$) and the sensitive cluster in the apparatus ($\alpha_2$) satisfies Conway and Kochen's \cite{Conway} definition of contextuality given  in the Introduction. It also answers an objection raised by Wigner against deterministic theories \cite{Wigner83}. Wigner considers a number of Stern-Gerlach apparatuses in series whose axes point alternately in the $z$ and in the $x$ direction, perpendicular to the direction $y$ of the particle entering the respective apparatus. Assume that the particle in the first apparatus escapes the contracting interaction in the screen with almost unchanged direction and with spin component in the +$z$ direction. Then the value of the determining hidden variable must lie in a fraction of the total range originally available for it. In the deflection in the subsequent apparatus, with axis in $x$ direction, the value must lie in a fraction of that fraction, and so on; so that after $N$ apparatuses, if $N$ is large enough, it would seem that the hidden variable lies in such a narrow range that it would determine the outcomes of all later experiments. This is in contradiction to the predictions of quantum mechanics. In the present proposal the pseudorandom phases of the wave functions representing the  clusters in the apparatuses ensure the continuing random appearance of the outcomes after any number  of apparatuses.

In traditional, indeterministic quantum mechanics, no superluminal signals can be transmitted because although the result of the sender can influence (steer) the result of the receiver, the sender cannot influence his own result, this being irreducibly indeterministic. In contrast to this, in the deterministic quantum mechanics here proposed, his result is determined by the form and the phase constant of the incoming wavepacket and the phase constants of all the grains in the measuring apparatus. Can he thus send superluminal signals? Notice that knowledge alone is not sufficient for this: whether or not he knows all the above-listed quantities of the wavepacket and the grains, the results and with them the correlations are the same. What would be required for communication is that he be capable of determining all those quantities. This is virtually impossible, signaling being anyway an operation between macroscopic bodies. We thus have determinism but no predictability,  just as in throwing dice.

The contextuality here introduced may also be taken as a concrete example of the
\vspace{-10pt}
\begin{quote}
impossibility of any sharp separation between the behaviour of atomic objects and the interaction with the measuring instruments which serve to define the conditions under which the phenomena appear,
\end{quote}
\vspace{-5pt}
as emphasized by Bohr \cite{Bohr49}.

\vspace{10pt}
\noindent
{\textbf{5~~Reproducing the Born rules}}
\smallskip

\noindent
In order to show that our conjecture is consistent with established laws, we here reproduce the Born rules; only here probabilities come in. We begin by defining what exactly we mean by the Born probability rules. We consider 3 cases:

$(\rmi)$ The probability of finding in a position measurement  a value lying in the interval $\rmd^3r$ about $\bfitr$ \cite[p.~19, 225, 226]{Cohen},  \cite[p.~117, 192]{Messiah}:
\[
\hspace{154pt}  P_1=|\psi(\bfitr)|^2\rmd^3 r, \hspace{140pt} (5.1)
\]
where  $\psi(\bfitr)$ is the incoming wavepacket  $\spsi{1}(\bfitr,t)$ of Section~4. We leave out the time variable (as in most textbooks). Here it is always the moment of contraction. 

$(\rmi \rmi)$
The probability of finding in a measurement  the eigenvalue $o_n$ of an observable with a \emph{discrete} non-degenerate spectrum \cite[p.~216]{Cohen},  \cite[p. 188-190]{Messiah}:
\[
\hspace{150pt} P_2=|(\psi_n(\bfitr),\psi(\bfitr))|^2, \hspace{119pt} (5.2)
\]
where $\psi_n(\bfitr)$ is the normalized eigenfunction associated with the eigenvalue $o_n$.
 
$(\rmi \rmi \rmi)$
The probability of finding an eigenvalue $a$ of an observable with a \emph{continuous} non-degenerate spectrum in the interval $\rmd a$ about $a$ \cite[p.~218]{Cohen},  \cite[p.~188, 189]{Messiah}:
\[
\hspace{147pt}  P_3= |(\psi(a;\bfitr),\psi(\bfitr))|^2 \rmd a, \hspace{105pt} (5.3)
\]
where $\psi(a;\bfitr)$ is the eigenfunction associated with the eigenvalue $a$.

\vspace{5pt}
It is important to note that in any case the Born probabilities only mean that a certain value \emph{is found}, not that the object measured \emph{has had} that value before the measurement. $P_1$ for example means that a position at \bfitr  \emph{ is found}, not that an object \emph{is} at {\bfitr}  when it is not observed. Generally, the probability of finding an object in a certain volume differs from the probability that the object is in that volume. If there is a needle in a haystack the probability of finding it depends on the amount of work and time spent in the search \cite{Bunge77}. Consider the normalization convention $\int_{\mathbb{R}^3} |\psi(\bfitr)|^2\rmd^3r$ =1. Of course, the probability that the needle, say,  \emph{is} somewhere must be 1, but the probability of \emph{finding} it somewhere may be considerably less than 1. The finding probability can only be 1 if the finding procedure  works with an \emph{efficiency} $\eta$ of 100\%. This shows generally that in the Copenhagen interpretation, formulas (5.1) to (5.3) refer to apparatuses which are tacitly presupposed to function in all cases with 100\%  efficiency. So we may put all constant efficiency factors, met in the derivation below, finally into the value of $\eta$, and then set $\eta=1$.

\vspace{5pt}
Now we are ready to deduce the Born probability rules in taking the absolute phase constants to be pseudorandom numbers. We do not deduce the Born rules from general fundamental principles \cite{Schlosshauer}, but reproduce them based on our conception of measurement and our criterion for contraction. That is, we are going to show how the
probability of recording a spot at the place of a phase-matching grain leads to the Born probability.

A good measuring apparatus must satisfy several conditions. In the ideal case the spots in the screen are small compared to the extensions of the individual eigenpackets and, in the case of discrete eigenvalues, also to the distance between them. The grain wavepackets in the screen differ only in their positions and phase constants. They produce a spot independent of the properties of the particular eigenpacket that covers them. Their centers $\bfitr_0$ are  distributed with constant density over the screen. To accomplish this or to correct for known deviations from it in the raw data, is the task of the experimenter.

We first consider the Born rule for position measurements. Like time, position in our theory is a parameter, not an observable (self-adjoint operator). Thus fanning out is not needed here. Owing to the described properties of the screen, the probability of phase matching (3.2a) in $\rmd^3r$ is independent of position $\bfitr$ and can be absorbed in the efficiency constant $\eta$. 

Given that (3.2a) is satisfied, the probability that the overlap condition (3.2b) is also satisfied and contraction occurs is just

\[
K:= \bigg[\int_{\mathbb{R}^3} |\spsi{1}(\bfitr)| \, |\spsi{2}(\bfitr)| \, {\textrm d}^3r\bigg]^2
\]
because the phase matching numbers $\alpha/2\pi$ are conceived to be uniformly distributed in the interval $[0,1]$, and any number which lies in the interval $[0,K]$ leads to contraction.  Thus the probability $P_4$ that a spot is produced in $\rmd^3r$ is
\[ P_4=\eta K\rmd^3r\,.\]
Due to the smallness of the volume  $\upsilon_{\rm gr}$  of the grain packet $\spsi{2}$ compared to  the volume of the incoming packet $\spsi{1}$, we expect that the integral $I$ in $K$
\[I:=
\int_{\mathbb{R}^3} |\spsi{1}(\bfitr)| \, |\spsi{2}(\bfitr)| \, {\textrm d}^3r
\]
can be approximated by $I=\eta\; |\psi_1(\bfitr_0)|$, where $\bfitr_0$ is the center of the grain packet $\psi_2$, and $\eta$ is independent of $\bfitr_0$ and the height of $\psi_2$ there. This is suggested by the fact that $I$ effectively extends only over the volume $\upsilon_{\rm gr}$ of the grain packet and we may apply the mean value theorem for integration in the form

\[
I:= |\spsi{1}(\bfitr_0^{\prime})| 
\int_{\upsilon_{\rm gr}} |\spsi{2}(\bfitr)| \, {\textrm d}^3r,
\]
where $\bfitr_0^{\prime}$ lies somewhere inside $\upsilon_{\rm gr}$. By the presupposed properties of the screen, the integral is the same for all grain packets, i.e. independent of $\bfitr_0$, so that $I=\eta  \, |\spsi{1}(\bfitr_0^{\prime})|$, with some constant $\eta$. And as $\upsilon_{\rm gr}$ is small, we may then approximate  $\bfitr_0^{\prime}$ by $\bfitr_0$, so that indeed, with some new $\eta$,
 \begin{displaymath}
\hspace{160pt}I = \eta \; \, |\spsi{1}(\bfitr_0)|. \hspace{140pt}(5.4)
\end{displaymath}
In Appendix B it is confirmed that any still existing dependence of this $\eta$ on the difference $\Delta=\bfitr_0^{\prime} - \bfitr_0$ is very weak, and that neglecting it is a fair approximation. With (5.4) and $K=I^2$ the probability $P_4$ turns into
\begin{displaymath}
P_4 = \eta^3 \,|\spsi{1}(\bfitr_0)|^2\, {\textrm d}^3r. 
\end{displaymath}

\noindent
Then setting $\eta=1$ and writing   $\bfitr$,  $\psi$ instead of  $\bfitr_0$,    $\spsi{1}$, respectively,  we arrive at
 \begin{displaymath}
 P_4=|\psi(\bfitr)|^2\rmd^3r,
 \end{displaymath}
 and this is the Born rule (5.1) for a position measurement.  
\vspace{5pt}

The Born rules for other than position measurements are obtained from the Born rule for a position measurement and the fanning out concept, independent of the special criterion for contraction. 

First consider an observable with \emph{discrete} eigenvalues $o_n$.
 Any spot observed in the spatial region $\Delta_n$, covered by the eigenfunction $\psi_n$, means that the eigenvalue $o_n$ has been observed. Thus the probability $P_5$ of observing $o_n$ is obtained by integrating the probability $P_4$ of observing a spot in $\rmd^3r$ over the region $\Delta_n$
\begin{displaymath}
P_5=\int_{\Delta_n}|\psi(\bfitr)|^2\,\rmd^3r= \int_{\Delta_n}\psi^*(\bfitr)\,\psi(\bfitr)\,\rmd^3r.
\end{displaymath}
In the second integral we use the expansion (4.1)
 \begin{displaymath}
 \hspace{151pt}\psi(\bfitr)=\sum_l c_l\,\psi_l(\bfitr) \hspace{132pt} (5.5)
 \end{displaymath}
 \vspace{-20pt}
 
 \noindent
 with $(4.2)$
 \vspace{-15pt}
 \begin{displaymath}
 \hspace{108pt}c_l=(\psi_l(\bfitr), \spsi{}(\bfitr))=\int \psi^*_l(\bfitr)\, \spsi{}(\bfitr)\, {\rm d}^3r.\hspace{85pt} (5.6)
 \end{displaymath}
But as the integral $\int_{\Delta_n}$ extends only over the region covered by $\psi_n$, and as the apparatus is presupposed to have enough resolving power, the sum $\sum_l$ can be replaced by one of its terms, so that (5.5) reduces to $\psi=c_n\,\psi_n$. With this we obtain
\begin{displaymath}
P_5=|c_n|^2\int_{\Delta_n}\psi_n^*(\bfitr)\,\psi_n(\bfitr)\,\rmd^3r,
\end{displaymath}
and as $\psi_n$ by definition is negligible outside $\Delta_n$, the integral may be extended over all space and is then 1 for the normalized $\psi_n$. With $c_n$ from (5.6) ($l\to n)$, we arrive at
\begin{displaymath}
\hspace{130pt}
P_5=|c_n|^2=|(\psi_n(\bfitr),\psi(\bfitr))|^2, \hspace{100pt} (5.7)
\end{displaymath}
and this is Born's rule (5.2) for a general discrete variable.

\vspace{5pt}
In the case of a \emph{continuous} spectrum no apparatus can have sufficient resolving power to separate the eigenvalues from each other, and we can only ask for the probability of observing an eigenvalue in some specified interval $\Delta a$ (cf. \cite[p.~260-265]{Cohen}). We begin by considering a \emph{discrete} spectrum whose eigenfunctions $\psi_n$ overlap in space. Having observed a spot in the interval $\Delta^3r$ then means that we have observed either $o_n$ or $o_{n+1}$ or ... . Thus the probability $P_6$ of observing an eigenvalue in the interval $\Delta^3r$ is the sum of the probabilities $P_5$ over the eigenvalues $o_n$ in the interval $\Delta_{\mbox{\it o}}$ that is associated with the interval $\Delta^3r$:
\[
P_6=\sum_{  \mbox{$o_n$}  \in \Delta_{\mbox{\it o}}  } |(\psi_n(\bfitr),\psi(\bfitr))|^2.
 \]
Now, the transition to a continuous spectrum consists in replacing the sum in $P_6$ by the integral (cf. [21, p. 217-218]):
\[
P_7=\int_{\Delta_{\mbox{\it a}}}|(\psi(a;\bfitr),\psi(\bfitr))|^2\,\rmd a,
\]
where $\psi(a;\bfitr)$ is the (improper) eigenfunction associated with the eigenvalue $a$ of a continuous spectrum.
From this it follows that the probability of observing an eigenvalue in the interval $\rmd a$ is
\[
\hspace{140pt}P_8=|(\;\psi(a;\bfitr),\psi(\bfitr)\,)|^2\,\rmd a, \hspace{100pt} (5.8)
\]
and we have arrived at Born's rule (5.3) for a general continuous spectrum.

A comparison of the transition probability $|(\psi_2|U\psi_1)|^2\;$   \cite[p. 725]{Messiah}     with our formulas (5.7) or (5.8) shows that $U\psi_1$ could be interpreted as the result of the fanning-out process.

\vspace{10pt}
The natural idea of a contracting wavepacket here introduced should be compared with the Copenhagen idea of a point particle inside the wavepacket, which, in order to account for the observed interference effects, is not allowed to exist at a definite position until a measurement is made. I hope the arguments brought forward in this article will break the evil spell cast on this part of quantum mechanics. This article, together with \cite{Jabs96}, \cite{Jabs15},  \cite{Jabs19},  \cite{Jabs13},  actually complete a consistent picture of the phenomena considered in quantum mechanics.

\vspace{15pt}
\noindent
{\textbf{Appendix A. Valuation of the absolute phase}}
\nopagebreak
\smallskip

\noindent
Here, we want to deal with the frequently encountered assertion that the absolute phase of a wave function is undetermined if the wave function represents a definite number of particles, in the same way as the position is undetermined if the wave function represents a particle of definite momentum \cite{Heitler} - \cite{Davydov}. This would contradict our approach, in which a definite phase is ascribed to every wave function. 

We plainly reject that assertion. It stems from the introduction of a phase operator, which does not commute with the particle-number operator. There are, however, serious difficulties with the construction of a phase operator. The original phase operator, introduced by Dirac \cite{Dirac27}, is not Hermitian and thus cannot be an observable. Subsequently there have been many attempts to construct a phase operator that is less deficient, but each proposal had its own difficulties and none has met with general approval. Details can be found in \cite{Mandel95} and in the literature cited there.

Moreover, in a plane wave $\exp(\rmi (\bfitk\bfitr -\omega t+\alpha))$, the phase constant appears on an equal footing with the time $t$. So, what holds for time should also hold for phase. Regarding time, Pauli \cite{Pauli58} pointed out 
that it is generally not possible to satisfy the canonical commutation relations between the operators time and energy; he wrote:
\begin{quote}
We, therefore, conclude that the introduction of an operator t is basically forbidden and the time t must necessarily be considered as an ordinary number (`c-number') in Quantum Mechanics.
\end{quote}

Indeed, there is a position operator but no time operator in non-relativistic quantum mechanics \cite[p.~252]{Cohen}, \cite{Pauli58} - \cite{Becker}. And in relativistic quantum field theory both position and time are parameters, not operators. We therefore take the stand that we do not need a phase operator for observing the phase, just as we do not need a time operator for observing the time. Dispensing with the phase operator frees us from the phase/number uncertainty relation and allows us to ascribe a definite phase to every wave function, though the actual values of the phases may be unknown to us.

\vspace{10pt}
\noindent
{\textbf{Appendix B. Approximation in Eq.~(5.4)}}
\smallskip

\noindent
We estimate here the degree of approximation in taking the $\eta$ in formula (5.4) as independent of the difference $\Delta =\bfitr_0^{\prime} -   \bfitr_0$,  under the conditions for a measurement of position. That is, we want to show that
\[
\int_{\mathbb{R}^3} |\spsi{1}(\bfitr)| \, |\spsi{2}(\bfitr)| \, {\textrm d}^3r
\approx\eta\; |\spsi{1}(\bfitr_0)|.
\] 
To this end we choose the measured packet $|\psi_1|^2$ and the grain packet $|\psi_2|^2$ to be one-dimensional (normalized) Gaussian functions with full width $\sigma_1$ and $\sigma_2$, respectively. Thus
\[
|\psi_1(x)|=\left(\frac{2}{\pi \sigma_1^2}\right)^{1/4}\, \exp\!\Big(-(x-x_0^{\prime})^2/\sigma_1^2\Big),
\]
\[
|\psi_2(x)|=\left(\frac{2}{\pi \sigma_2^2}\right)^{1/4}\, \exp\!\Big(-(x-x_0)^2/\sigma_2^2\Big).
\]
The center of the grain function $|\psi_2|$ lies at the position $x_0$, while the center of the measured function $|\psi_1|$ lies at $x_0^{\prime}$. The value of the integral
\[I:=
\int_{-\infty}^{+\infty} |\spsi{1}(x)| \, |\spsi{2}(x)| \, {\textrm d}x
\]
can then be evaluated exactly, with the result
\[\hspace{48pt}
I(\delta)=\left[\frac{2}{\sigma_1/\sigma_2+\sigma_2/\sigma_1}\right]^{1/2}\,\exp\!\left(\frac{\delta^2}{\sigma_1^2}\bigg[\frac{1}{1+(\sigma_1/\sigma_2)^2}-1\bigg]\right), \hspace{50pt} (\rm B1)
\]
\vspace{10pt}
where $\delta=x_0-x_0^{\prime}$. This can be written as

$
 \hspace{100pt}I(\delta)=Q(\delta)\,\times |\psi_1(x_0)|  \hspace{20pt} \textrm{with}
$
\[\hspace{38pt}
Q(\delta)=\left[\frac{2}{\sigma_1/\sigma_2+\sigma_2/\sigma_1}\right]^{1/2}\,  \bigg(\frac{\pi\sigma_1^2}{2}\bigg)^{\!1/4}
\exp\!\left(\frac{\delta^2}{\sigma_1^2}\bigg[\frac{1}{1+(\sigma_1/\sigma_2)^2}\bigg]\right). \hspace{25pt} (\rm B2)
\]

\noindent
From (B1) and (\rm B2) it is seen that $Q(\delta)$ varies only slowly with $\delta$ where $I(\delta)$ has appreciable values. For example, with $\sigma_2=0.05\sigma_1$ (corresponding to the small  $v_{\rm {gr}}$), the value of $Q$ at $\delta/\sigma_1=2$ has increased by only 1\% from its value at $\delta/\sigma_1=0$, whereas $I$ thereby has decreased to 2\% of its value at $\delta=0$, and with this rarely satisfies the overlap condition (3.2b). We therefore conclude that (5.4) is a fair approximation.

\vspace{30pt}

\noindent
{\textbf{Notes and references}}
\renewcommand{\labelenumi}{[\arabic{enumi}]}
\renewcommand{\section}[2]

\bigskip
\hspace{5cm}
------------------------------

\begin{thebibliography}{99}
\bibitem{note2} In the theory of Ghirardi, Rimini and Weber \cite{Ghirardi}, \cite[Ch.~22]{Bell87}, \cite{Bassi} the reduction (collapse) is also a contraction to a Gaussian function,  here chosen ad hoc, whose dimension is not, however, determined by the physical objects involved but by a new constant of nature of the order of $10^{-7}$ m. Moreover, like the other collapse theories \cite{Pearle}, \cite{Brusheim}, that theory is not deterministic.
\end{thebibliography}
\end{document}